# Long-term propagation of $CO_2$ plume below the seal accounting for chemical reactions and water counter-flow

Amin Shokrollahi, Syeda Sara Mobasher, Kofi Ohemeng Kyei Prempeh, Sara Borazjani, Abbas Zeinijahromi, Pavel Bedrikovetsky

*School of Chemical Engineering, Discipline of Mining and Petroleum Engineering, The University of Adelaide, Adelaide, SA 5005, Australia*

**Abstract**

Accurate prediction of long-term $CO_2$ plume migration beneath seals is crucial for the viability of $CO_2$ storage in deep saline aquifers. Groundwater counterflow and chemical reactions between $CO_2$, brine, and rock significantly influence plume dynamics, dispersion, and boundary evolution. This study develops a novel analytical framework by deriving governing equations for gravity-driven gas migration and obtaining exact self-similar solutions for both pulse and continuous injection. The approach explicitly incorporates chemical reactions and water counterflow, addressing factors often neglected in existing analytical models. A sensitivity analysis is conducted to evaluate the effects of key parameters on $CO_2$ migration. The results show that increasing the chemical reaction coefficient delays the first arrival of $CO_2$ while accelerating its final arrival, highlighting the impact of geochemical reactions on plume migration. Additionally, higher up-dip water velocity shortens the accumulation period, thereby reducing the time required for $CO_2$ to become trapped in geological formations. These findings improve predictive capabilities for long-term $CO_2$ storage and provide insights for optimising sequestration strategies.

# Nomenclature

**Parameters**

| | |
|---|---|
| $A(h)$ | Stringer cross-section |
| $A(h_L)$ | Stringer cross-section for typical current thickness |
| $A_D$ | Dimensionless stringer cross-section |
| $C(h)$ | Integral of $A(h)$ |
| $G$ | Constant for continuous injection |
| $g$ | Gravitational acceleration constant |
| $H$ | Vertical height of reference point |
| $h(x,t)$ | Current thickness |
| $h_D$ | Dimensionless current thickness |
| $h_L$ | Reference thickness of gravity current |
| $K_a$ | Chemical reaction coefficient |
| $K_{rgcw}$ | Gas relative permeability in presence of connate water |
| $k$ | Permeability |
| $k_a$ | Dimensionless chemical reaction coefficient |
| $L$ | Direct distance between the trap and the starting point of hydrocarbon migration below the seal |
| $l$ | Power of power-law shape of stringer |
| $M$ | Mass of expulsed gas |
| $m, n$ | Powers in the self-similar solution |
| $p$ | Pressure |
| $p_H$ | Pressure of the reference point on z-axis |
| $Q$ | Gas injection rate |
| $R$ | Equilibrium gas concentration in water |
| $r$ | Defined power as a function of $l$ |
| $s_{cw}$ | Connate water saturation |
| $s_{gr}$ | Residual gas saturation |
| $t$ | Time |
| $T$ | Injection period during the pulse injection |
| $t_D$ | Dimensionless time |
| $U$ | Gas velocity/flux |
| $u$ | Gas buoyant velocity |
| $W$ | Module of water velocity |
| $w$ | water-assisted gas velocity |
| $x$ | Coordinate along the seal |
| $x_D$ | Dimensionless coordinate along the seal |
| $\alpha$ | Angle between seal to horizontal axes |
| $\beta$ | Constant of power-law shape of stringer |
| $\tau$ | New time scale |
| $\varepsilon$ | Gravity-diffusion number |
| $\mu_g$ | Gas viscosity |
| $\mu_w$ | Water viscosity |
| $\rho_w$ | Water density |
| $\Delta\rho$ | Density difference between water and gas |

# 1  Introduction

Gravity currents in porous media occur when one fluid spreads through a domain that is saturated with a different fluid. The balance between pressure and buoyancy is maintained by viscous forces within the pore space (Longo et al., 2015). This phenomenon is central to various natural and industrial processes, such as hydrocarbon migration in geological basins, $CO_2$ and hydrogen storage in geological formations (Dentz and Tartakovsky, 2009; Williams et al., 2018), gas injection into depleted oil fields, groundwater flow (Fahs et al., 2014), and the movement of dense non-aqueous phase liquid contaminants in subsurface porous layers (Tosco et al., 2014).

A detailed understanding of gravity currents is essential for effectively modelling $CO_2$ storage in underground reservoirs. This understanding helps optimise $CO_2$ injection strategies and ensures the safe and efficient containment of $CO_2$ over long periods. Research into gravity currents in simple, homogeneous media with basic geometries has been extensive, particularly in horizontal and inclined beds (Huppert and Woods, 1995; Lyle et al., 2005; Vella and Huppert, 2006). However, real-world applications, such as $CO_2$ storage or hydrocarbon migration in geological formations, are more complex and require consideration of factors such as topography, medium heterogeneity, capillary pressure, chemical reactions, natural water flux, and seal asperities.

The influence of topography on gravity currents has been widely studied. Golding and Huppert (2010) conducted both theoretical and experimental research to explore the effect of confining boundaries on gravity currents in porous media. They found that in one-dimensional gravity-driven flows through porous channels of uniform cross-section, the shape of the channel significantly affects the propagation rate. The time exponent related to the current volume is influenced by the geometry of the channel. However, they also showed that the confining boundaries have little effect on the propagation rate when the slope of the channel is steeper than the slope of the free surface. This finding is particularly relevant for $CO_2$ storage, as the shape of the geological formations, including variations in slope, can significantly impact the migration of $CO_2$ in the subsurface.

Further studies, such as those by Pegler et al. (2013), examined the effect of upward-sloping topography on gravity currents in two- and three-dimensional porous media. They concluded that topography plays an important role in the evolution of gravity currents, particularly during the early and late stages of current propagation. The shape of the lower boundary influences the behaviour of the current, with different topographies leading to distinct flow patterns. These insights are crucial for understanding how $CO_2$ behaves when injected into subsurface formations with varying topography.

Zheng et al. (2015) also studied the dynamics of fluid flow in confined porous media, considering both theoretical and numerical approaches. Their research focused on the injection of one fluid into a porous medium saturated with another fluid of different density and viscosity. In the early stages of fluid flow, buoyancy primarily drives the movement, leading to a nonlinear diffusion equation with a self-similar solution. In contrast, later stages are driven by injection pressure, leading to a nonlinear hyperbolic equation that determines the propagation rate. This transition from pressure -driven to buoyancy-driven flow is highly relevant for $CO_2$ storage, as $CO_2$ injection often begins with pressure -driven migration, which later shifts to buoyancy-driven flow as the injection progresses.

Medium heterogeneity is another important factor that influences gravity currents. Ciriello et al. (2013) developed a new formulation to assess the impact of permeability variations in both vertical and horizontal directions on the propagation of gravity currents. By using power-law functions to model permeability variations, they derived a self-similar solution for current propagation. Their work highlighted the critical role of permeability heterogeneity in altering the behaviour of gravity currents, which is essential for accurately modelling $CO_2$ storage. Variations in permeability can significantly affect the movement of $CO_2$ through the reservoir, potentially influencing its migration and trapping.

Zheng et al. (2013, 2014) also investigated the impact of porosity and permeability gradients, both parallel and transverse to the flow direction, on gravity currents. They showed that these heterogeneities result in different types of self-similar solutions and affect the shape and extent of the current. The presence of macro-heterogeneity, or large-scale variations in the medium's properties, can substantially influence the dynamics of gravity currents. For $CO_2$ storage, such heterogeneities are important because they can affect how $CO_2$ is distributed within the reservoir, potentially influencing both storage capacity and containment.

Regarding the effect of chemical reactions on gravity currents, particularly during $CO_2$ storage in aquifers, these reactions play a significant role in influencing the movement of $CO_2$ within the reservoir. After $CO_2$ injection, it initially exists as a free-phase fluid within the reservoir. Over time, it dissolves into the formation water, triggering a range of geochemical reactions. Some of these reactions may be advantageous, as they can contribute to the long-term containment of $CO_2$ by converting it into dissolved forms or by forming new carbonate minerals, thus reducing its mobility. On the other hand, certain reactions could negatively impact the storage process by facilitating the migration of $CO_2$, either by altering the permeability of the reservoir or by creating new pathways that allow for the gas to move more easily (Carroll et al., 2016; Rochelle et al., 2004).

These chemical processes directly influence the dynamics of gravity currents, which govern the spread of $CO_2$ within the reservoir. The extent to which $CO_2$ is trapped or allowed to migrate depends on the specific mineralogical, structural, and hydrogeological properties of the local geological formations. Therefore, it is essential for each $CO_2$ storage operation to account for the unique geological and chemical conditions at the site, as these will determine how the chemical reactions interact with the gravity currents and ultimately affect the behaviour and $CO_2$ storage capacity.

This study develops dynamic equations for gravity currents and derives their self-similar solutions to characterise the long-term evolution of $CO_2$ plumes under different injection scenarios. The solution for pulse injection, applicable when the injection period is shorter than the migration period within the reservoir, provides insights into plume behaviour over extended timescales. In contrast, the solution for continuous injection, relevant when the injection period exceeds the migration period, introduces a time-dependent scaling factor that influences long-term propagation. A key advancement of this work is the derivation of exact solutions for plume dynamics under both injection regimes, explicitly accounting for chemical reactions and water counterflow beneath the moving plume. These analytical solutions enable a more rigorous assessment of $CO_2$ migration and storage behaviour, enhancing predictive capabilities for long-term sequestration in deep saline aquifers. Additionally, analytical simulations and sensitivity analyses are conducted to evaluate the influence of key parameters, including the chemical reaction coefficient and brine counterflow velocity, on plume migration. The findings contribute to a deeper understanding of the factors governing $CO_2$ storage efficiency and provide a foundation for optimising sequestration strategies in geological formations.

The manuscript is divided into the following sections. Section 2 presents the derivation of the governing equations, accounting for chemical reactions and water counter-flow, along with the exact solutions for both pulse and continuous injection cases. Section 3 discusses the results of a multivariate sensitivity analysis and includes a practical example of $CO_2$ injection. Section 4 explores the extensions, limitations, and practical applications of the analytical model. Section 5 concludes the study.

# 2  Governing Equations

This section outlines the mathematical model for $CO_2$ injection in reservoirs. It begins with the assumptions underlying the model (Section 2.1). The governing equations for buoyancy-driven segregated two-phase flow are then presented (Section 2.2). Following this, the dimensionless form of the advection-diffusion-reaction equations is derived (Section 2.3). Finally, Section 2.4 presents the exact solutions for pulse injection, which describes the long-term evolution of the $CO_2$ plume, and continuous injection, which introduces a time-scale that affects the long-term propagation of the plume.

## 2.1  Assumptions of the model

This study presents an analysis of the buoyancy-driven two-phase migration of injected $CO_2$ through sealed pathways. Figure 1 depicts the inclined top of the geological formation, where gas injection takes place beneath the seal, illustrating the buoyancy-driven migration of gas along a tortuous path within the reservoir. Different colours represent the gas plume shape at various times, from injection to entrapment within the trap. The solid navy curve represents the envelope curve, which indicates the height of the gas plume at any given time and position. *U* denotes the gas velocity driven by buoyancy, and *W* corresponds to the water velocity beneath the advancing gas plume, and $h(x,t)$ represents the gas plume height at a given time and position.

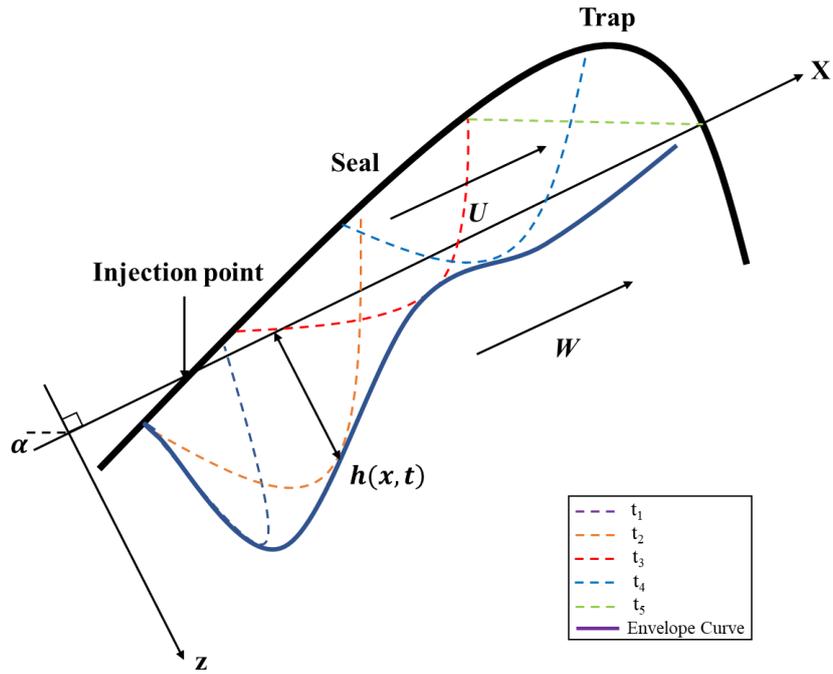

Figure 1. Buoyancy-driven segregated flow of injected $CO_2$ below the seal.

The model is developed based on a set of assumptions that simplify the analysis of $CO_2$ migration within the reservoir. First, both gas and water are treated as incompressible fluids, assuming constant fluid densities throughout the process. In the carrier bed beneath the gas plume, it is assumed that water flows in a steady-state manner, parallel to the seal, during both the gas injection and buoyant propagation phases. Furthermore, the water pressure in the flow zone beneath the gas–water contact (GWC) is assumed to remain unaffected by the gas injection, ensuring that it is not disturbed during the migration of the gas. The model also accounts for chemical reactions between the gas, water, and rock, incorporating a chemical reaction constant that varies depending on the system properties. These reactions can influence $CO_2$ solubility and contribute to the formation of minerals within the reservoir. Finally, flow perpendicular to the seal in both the gas and water zones is neglected, allowing for a hydrostatic pressure distribution across the flow cross-section. These assumptions provide a structured and simplified basis for deriving the governing equations and solutions that describe $CO_2$ migration in the reservoir.

## 2.2 Governing Equations for Buoyancy Driven Segregated Two-Phase Flow

The governing equation for buoyancy-driven segregated two-phase flow of incompressible fluids in porous media is represented by a system of two partial differential equations. This system includes the mass balance equation and Darcy's law, which together describe the evolution of the hydrocarbon plume thickness, $h(x,t)$, over time.

In Figure 1, the $x$-axis represents the direction of the seal, while the z-axis is perpendicular to it. The parameter $\alpha$ denotes the slope of the inclined layer, and the thickness of the gas layer is represented by $h(x,t)$. The velocity of the water is described by Darcy's law, as follows:

$$W = -\frac{k}{\mu}\left(\frac{\partial p}{\partial x} + \rho_w g \sin\alpha\right) \tag{1}$$

Here, $W$ denotes the water velocity, $k$ is the permeability, $\rho_w$ represents the water density, $\mu_w$ is the water viscosity, and $g$ is the gravitational acceleration. Using the pressure gradient obtained from equation (1), the pressure in the gas zone, relative to any reference point on the $z$-axis with a reference pressure $p_H$, can be expressed as follows:

$$0 < z < h(x,t):$$
$$p(x,z) = p_H - \rho_w g \cos\alpha (H\cos\alpha) - \left(\frac{\mu_w}{k}W + \rho_w g \sin\alpha\right)x + \Delta\rho g h \cos\alpha + \rho_g gz \cos\alpha \tag{2}$$

Here, $H$ denotes the corresponding vertical height of the reference point, and $\Delta\rho$ is the density difference between water and gas. Using equation (2), the pressure gradient along the $x$-axis in the moving gas zone is given by:

$$\frac{\partial p}{\partial x} = -\left(\frac{\mu_w}{k}W + \rho_w g \sin\alpha\right) + \Delta\rho g \cos\alpha \frac{\partial h}{\partial x} \tag{3}$$

Substituting equation (3) into Darcy's law for gas and introducing the following reference velocities results in:

$$w = K_{rgcw} \frac{\mu_w}{\mu_g} W, \quad u = \frac{kK_{rgcw} \Delta \rho g \sin \alpha}{\mu_g} \tag{4}$$

yields the following equation for the gas velocity:

$$U = \left\{ w + u - uctg\alpha \frac{\partial h}{\partial x} \right\} \tag{5}$$

Here, $U$ represents the moving gas velocity, $u$ is the buoyant velocity, $w$ denotes the water-assisted gas velocity, and $K_{rgcw}$ is the gas relative permeability in the presence of connate water. The mass balance equation for the moving gas plume below the seal, accounting for gas incompressibility and incorporating chemical reactions in the sink term, is given by:

$$\phi S_g \frac{\partial A(h)}{\partial t} + \frac{\partial A(h)U}{\partial x} = -K_a A(h)$$
$$S_g = 1 - S_{cw} + RS_{cw} - s_{gr} \left( \frac{\partial h}{\partial t} \right) \tag{6}$$

where $A(h)$ is the stringer cross-sectional area, $R$ is the equilibrium gas concentration in water, $\phi$ is the porosity, $S_g$ is the gas saturation, $K_a$ is the chemical reaction coefficient, $s_{cw}$ is the connate water saturation, and $s_{gr}$ is the residual gas saturation.

Substituting the migration flux expression (5) into the general mass balance equation (6) for the gas plume yields the following advective-diffusive equation:

$$\phi S_g \frac{\partial A(h)}{\partial t} + (w+u) \frac{\partial A(h)}{\partial x} = uctg\alpha \frac{\partial}{\partial x} \left[ A(h) \frac{\partial h}{\partial x} \right] - K_a A(h) \tag{7}$$

where $A(h)$ is stringer cross-section, $\phi$ is porosity, $K_a$ is chemical reaction coefficient, $S_g$ is gas saturation.

2.3 Dimensionless advection-diffusion-reaction equation

Using the following dimensionless variables and parameters:

$$x_D = \frac{x}{L}, \; t_D = \frac{(w+u)t}{L\phi S_g}, \; h_D = \frac{h}{h_L}, \; A_D = \frac{A(h)}{A(h_L)}, \; \varepsilon = \frac{uctg\alpha h_L}{(w+u)L}, \; k_a = \frac{K_a L}{(w+u)}$$

$$C'(h_D) = A_D(h_D), \; C(h_D) = \int_0^{h_D} A_D(v)dv$$

(8)

Equation (7) can be simplified as follows:

$$\frac{\partial A_D}{\partial t_D} + \frac{\partial A_D}{\partial x_D} = \varepsilon \frac{\partial^2 C(A_D)}{\partial x_D^2} - k_a A_D \tag{9}$$

where $h_L$ is a typical current thickness and $L$ is the direct distance between the trap and the starting point of hydrocarbon migration below the seal and $\varepsilon$ is the gravity-diffusion number. To simplify the solution of the partial differential equation (9), the following transformation is applied, which converts the reference system into a unitary speed system using the following transformation:

$$A_D(x_D, t_D) \to A_D(z, t_D), \; z = x_D - t_D \tag{10}$$

Applying transformation (10), the partial differential equation (9) becomes:

$$\frac{\partial A_D}{\partial t_D} = \varepsilon \frac{\partial}{\partial z}\left[\frac{\partial C(A_D)}{\partial z}\right] - k_a A_D \tag{11}$$

## 2.4 Analytical solutions

The purpose of this study is to model gas propagation as distinct gravity streams or currents within stringers. To account for the asperities of the seal, the cross-section of the seal stringer is assumed to have a power-law shape, expressed by the following formula:

$$A_D(h_D) = \beta h_D^l \tag{12}$$

The $l$ and $\beta$ in this equation are the constants that determine the shape of the cross-section. Figure 2 presents various forms of power-law stringer cross-sections for varying $l$ values. As illustrated, increasing $l$ leads to a larger cross-sectional area of the stringers, which in turn reduces the velocity of migrating gas for a constant volumetric flow rate within the stringers.

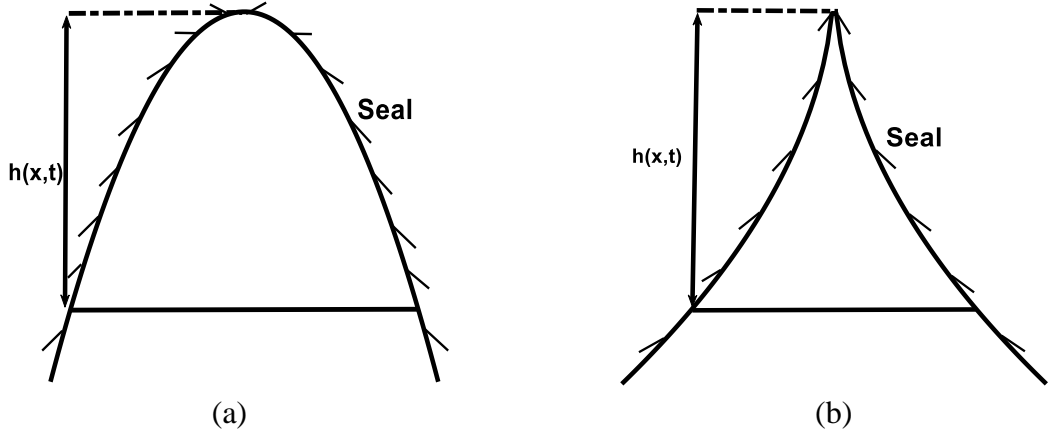

(a)                  (b)

Figure 2. Stringers cross-sections for (a) *l>1* and (b) *l<1*.

Accounting for the power-law expression (12) of the stringer cross-section area and transforming the unknown function as follows:

$$A(h_D) = e^{-k_a t_D} a(h_D), \quad c(h_D) = \int_0^{h_D} a(h)\, dh \tag{13}$$

Equation (11) becomes:

$$\frac{\partial a}{\partial t_D} = \beta^{-\frac{1}{l}} \varepsilon \frac{e^{-k_a t_D / l}}{l+1} \frac{\partial^2}{\partial z^2}\left(a^{\frac{l+1}{l}}\right) \tag{14}$$

Changing time scale from $t_D$ to $\tau$ and introducing power $r$

$$\tau = \beta^{-\frac{1}{l}} \frac{\varepsilon}{k_a r}\left(1 - e^{-k_a t_D / l}\right), \quad r = \frac{l+1}{l} \tag{15}$$

transforms equation (14) to the form

$$\frac{\partial a}{\partial \tau} = \frac{\partial^2}{\partial z^2}\left(a^r\right) \tag{16}$$

To solve the partial differential equation (16), a self-similar solution is employed in this study (Barenblatt, 2003; Bedrikovetsky and Borazjani, 2022). The following self-similar form of the solution is considered:

$$a(z,\tau) = \frac{1}{\tau^m} f\left(\frac{z}{\tau^n}\right) \tag{17}$$

Substituting the form (17) into equation (16) yields:

$$-\frac{1-2n}{r-1}f(y)-nyf'(y)=(f^r)'', \quad y=\frac{z}{\tau^n}, \quad m=\frac{1-2n}{r-1} \tag{18}$$

To solve the ordinary differential equation (18), initial and boundary conditions must be applied. Once this equation is solved, the final solution is obtained by substituting the calculated values of $m$, $n$, and the function $f$ into equation (17).

In this study, two injection scenarios, pulse injection and continuous injection, are examined. Pulse injection refers to cases where the injection period is shorter than the propagation time of $CO_2$ beneath the seal. The solution for pulse injection describes the long-term evolution of the plume when the injection duration is shorter than the $CO_2$ migration period in subsurface reservoirs. Continuous injection, on the other hand, occurs when the injection period exceeds the migration period of $CO_2$ in subsurface reservoirs. This can take place during the early stages of injection. Additionally, using the solution for continuous injection introduces a time scale into long-term propagation.

The following section presents the analytical solutions for both pulse and continuous injection scenarios.

### 2.4.1 Pulse Injection

The analytical solution for the spreading of instantly injected gas beneath the dipping seal is presented in this section. Since the stringers cannot extend to infinity at final times, the following boundary conditions can be formulated:

$$x \to \infty: h=0; \quad x \to -\infty: h=0 \tag{19}$$

The mass $M$ of gas that was initially injected results in the following initial conditions:

$$t=0, \; x \notin 0: h(x,0)=0, \; A(x,0)=0, \; a(x,0)=0$$
$$\int_{-\infty}^{\infty} A(x,0)dx = \frac{M}{\phi \rho_g S_g} \tag{20}$$

Applying the dimensionless variable (8) transforms the initial boundary condition (20) into:

$$\int_{-\infty}^{\infty} A_D(x_D, 0) dx_D = \frac{M}{\phi \rho_g S_g A(h_L) L} = 1 \tag{21}$$

Applying transformations (10) and (13) to (21) leads to:

$$\int_{-\infty}^{\infty} a(z, 0) dz = 1 \tag{22}$$

The following relations for the constants *m* and *n* can be obtained by applying the self-similar form (17) to equation (22):

$$\int_{-\infty}^{\infty} \frac{\tau^n}{\tau^m} f\left(\frac{z}{\tau^n}\right) d\left(\frac{z}{\tau^n}\right) = \int_{-\infty}^{\infty} f(y) dy = 1, \, m = n \tag{23}$$

The following values for *m* and *n* can be obtained by solving equations (18) and (23) simultaneously:

$$\begin{cases} m = \dfrac{1-2n}{r-1} \\ m = n \end{cases} \rightarrow m = n = \frac{1}{r+1} \tag{24}$$

Using the values of *m* and *n*, the ordinary differential equation (24) becomes:

$$\frac{1}{r+1}(yf)' = (f^r)'' \tag{25}$$

Solving equation (25) subject to the conditions (19) and (22), yields the dimensionless exact solutions (26) and (27) for the gas front trajectory and the height of the injected gas in a pulse injection scenario, respectively.

$$x_D = t_D \pm \frac{\left(2\beta^{\frac{-1}{l}} \dfrac{\varepsilon(r+1)}{k_a(r-1)}\left(1 - e^{-k_a t_D / l}\right)\right)^{\frac{1}{r+1}}}{\left(2\int_0^{\pi/2} [\cos \delta]^{\frac{r+1}{r-1}} d\delta\right)^{\left(\frac{r-1}{r+1}\right)}} \tag{26}$$

Where $x_D$ is the dimensionless position of the front, $t_D$ is the dimensionless time, $\varepsilon$ is the gravity-diffusion number, $k_a$ is the dimensionless chemical reaction coefficient, and *r* is defined as a power function of *l* presented in equation (15). Equation (26) calculates the position of the initial and rear fronts using the '+' and '−' signs, respectively. The dimensionless height of moving gas plume can be calculated using equation (27).

$$h_D(x_D, t_D) = \frac{e^{-(k_a t_D(1-r))}}{\left(\beta^2 \frac{\varepsilon}{k_a r}\left(1-e^{-k_a t_D/l}\right)\right)^{\frac{k-1}{k+1}}} \left[\left(\frac{\sqrt{\frac{(r-1)}{2r(r+1)}}}{2\int_0^{\pi/2}[\cos\delta]^{\frac{r+1}{r-1}}d\delta}\right)^{\frac{2(r-1)}{r+1}} - \left(\frac{(r-1)}{2r(r+1)}\right)\left(\frac{x_D - t_D}{\left(\beta^{\frac{-1}{l}} \frac{\varepsilon}{k_a r}\left(1-e^{-k_a t_D/l}\right)\right)^{\frac{1}{r+1}}}\right)^2\right] \quad (27)$$

### 2.4.2 Continuous Injection

The analytical solution for the spreading of continuously injected gas beneath the dipping seal is presented in this section. For continuous injection, the boundary conditions remain the same as those for pulse injection, as defined in equation (19). In this scenario, the mass of gas in the system increases over time and can be expressed as follows:

$$\int a\,dz = G\tau, \quad G = \frac{\beta^{\frac{1}{l}} Q l k}{\varepsilon(w+u)A(h_L)}, \quad G = \text{constant} \quad (28)$$

Where *Q* is the rate of gas injection. To determine the values of *m* and *n*, the self-similar form (17) is substituted into the total mass equation (28) as follows:

$$\int \frac{\tau^n}{\tau^m} f(\frac{z}{\tau^n})d(\frac{z}{\tau^n}) = G\tau \rightarrow \int f(y)dy = G\tau \rightarrow m - n = -1 \quad (29)$$

The following values can be obtained for *m* and *n* using (18) and (29):

$$\begin{cases} m = \frac{1-2n}{r-1} \\ m - n = -1 \end{cases} \Rightarrow \begin{cases} n = \frac{r}{r+1} \\ m = \frac{-1}{r+1} \end{cases} \quad (30)$$

Substituting the values of *m* and *n* into the ordinary differential equation (18) results in the following ordinary differential equation:

$$f'' = \frac{1}{r(r+1)} f^{(2-r)} - \frac{1}{(r+1)} y f^{(1-r)} f' - (r-1) f^{(-1)} f'^{(2)} \tag{31}$$

Equation (31) does not have an exact solution and must be solved numerically. To solve this equation, the following condition must be considered:

$$\left.\frac{df}{dy}\right|_{y=0} = -\frac{G}{2rf^{(r-1)}(0)} \tag{32}$$

Using condition (32), the following dimensionless equation can be obtained for the calculation of the gas front in continuous Injection:

$$x_D = t_D \pm y_0 \left( \beta^{\frac{-1}{l}} \frac{\varepsilon}{k_a r} \left(1 - e^{-k_a t_D / l}\right) \right)^{\frac{r}{r+1}}, \quad f(y_0) = 0 \tag{33}$$

Equation (33) determines the positions of the initial and rear fronts using the '+' and '−' signs, respectively. This equation does not have an exact solution and must be solved numerically. The value of $y_0$ in Equation (33) corresponds to the point where the solution of Equation (31) equals zero. Figure 3 presents a schematic of the numerical solution to Equation (31) and illustrates the position of $y_0$. Also, the dimensionless height of moving gas plume can be calculated using equation (34).

$$h_D(x_D, t_D) = \frac{e^{-(k_a t_D (1-r))}}{\left( \beta^{-2r} \frac{\varepsilon}{k_a r} \left(1 - e^{-k_a t_D / l}\right) \right)^{\frac{1-r}{r+1}}} \left[ f\left( \frac{x_D - t_D}{\left( \beta^{\frac{-1}{l}} \frac{\varepsilon}{k_a r} \left(1 - e^{-k_a t_D / l}\right) \right)^{\frac{r}{r+1}}} \right) \right]^{r-1} \tag{34}$$

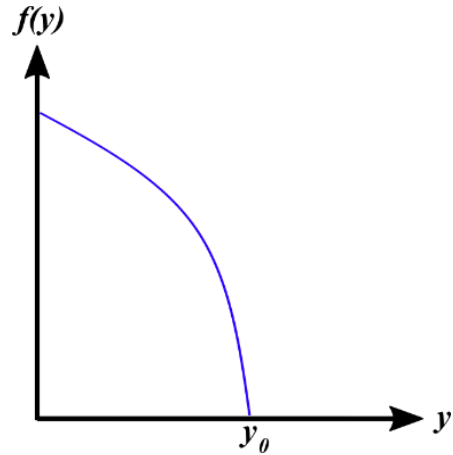

Figure 3. Schematic of the numerical solution of equation (31) and the position of $y_0$.

## 3 Sensitivity Analysis

A sensitivity analysis is performed to examine the influence of several parameters, including the chemical reaction coefficient ($k_a$) and the water counter-flow velocity ($w$) on the trajectory of the gas front and its arrival times.

To carry out a comprehensive and realistic sensitivity analysis, a basic example of $CO_2$ storage is considered, reflecting typical reservoir conditions. In this scenario, the reservoir has a width of 100 m and a height of 20 m. The trap is located 20 km from the injection point, with the seal inclined at an angle of 15 degrees. The reservoir's permeability is set at 10 mD, with a porosity of 0.1, and the depth of the reservoir is 10 km. The injection well has a radius of 0.1 m, and the velocity of the injected gas at the well wall is assumed to be $10^{-3}$ m/s and the injection time of 10 years.

Using this configuration, the sensitivity analysis is performed to assess the impact of the previously mentioned parameters on the front trajectory and the arrival times of the $CO_2$ plume. This approach provides valuable insights into the relative significance of each parameter under realistic reservoir conditions.

Figure 4 illustrates the position of the $CO_2$ plume at various times throughout the injection process. The highest point on the graph corresponds to the early stages of injection, where the plume is more concentrated and confined. Over time, as the injection progresses, the plume spreads beneath the seal, and the height of the plume decreases. Each curve in Figure 4 represents the position and shape of the $CO_2$ plume at different time intervals. Notably, each curve is characterised by two distinct fronts: the right front, which represents the initial front of the plume, and the left front, known as the rear front.

The arrival time is defined as the moment when the initial front reaches the trap, as indicated by the green dashed curve in Figure 4. The last arrival corresponds to the time at which the rear front reaches the trap, as indicated by the dashed purple curve in Figure 4. The difference between the first and last arrival times is referred to as the accumulation period, which represents the duration it takes for the injected $CO_2$ plume to be fully contained within a particular trap. This accumulation period is a crucial metric for evaluating the efficiency and long-term stability of $CO_2$ storage, as it provides insight into the timescale over which the plume migrates and becomes stabilised within the reservoir. By examining the evolution of the plume's position and shape over time, it is possible to assess the dynamics of $CO_2$ storage, including the impact of various geological and operational factors on plume migration.

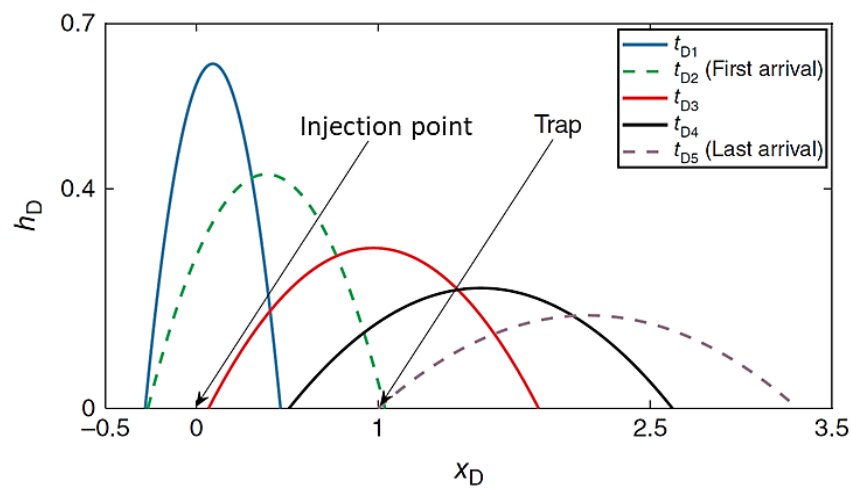

Figure 4. Representation of the first and last arrival times using the obtained analytical solution.

The following sections present a sensitivity analysis of the key parameters involved in the analytical model. In this section, the sensitivity analysis is conducted using the pulse injection solution; however, the results from the continuous injection solution follow the same trend as those of the pulse injection.

### 3.1 Effect of water counter-flow on $CO_2$ plume behaviour

One of the significant parameters influencing the behaviour of the migrating $CO_2$ plume following injection is the counter-flow of water beneath the moving plume. When water flows counter to $CO_2$ injection, it slows the plume's advance by restricting its movement through the pore spaces. Conversely, co-flowing water can enhance displacement efficiency. Changes in pressure and saturation due to water movement may accelerate or decelerate $CO_2$ migration, depending on the flow conditions. Additionally, water flow can induce fractures or preferential pathways, altering the migration pattern. The shape of the $CO_2$ plume is also affected, with counter-flowing water spreading it more widely and reducing storage efficiency.

To investigate the effect of water counter-flow on $CO_2$ plume behaviour, the model developed for pulse injection is applied. To ensure a rigorous analysis, the water velocity ($w$) value is obtained from the literature (Bjørlykke 1993; Hantschel and Kauerauf, 2009; Elizabeth and Valiya, 2014). According to these sources, water velocity ranges from $10^{-6}$ to $10^{-12}$ m/s; however, most references report a value of $10^{-9}$ m/s.

Figure 5 illustrates the effect of water velocity on the first and last arrival times of $CO_2$ at the trap. As shown in the figure, for a fixed $l$ value, an increase in upward velocity reduces the accumulation period, defined as the difference between the first arrival time (dashed curves) and the last arrival time (solid curves). This supports the underlying physics that upward water movement helps the $CO_2$ plume to advance faster and be trapped in a shorter time.

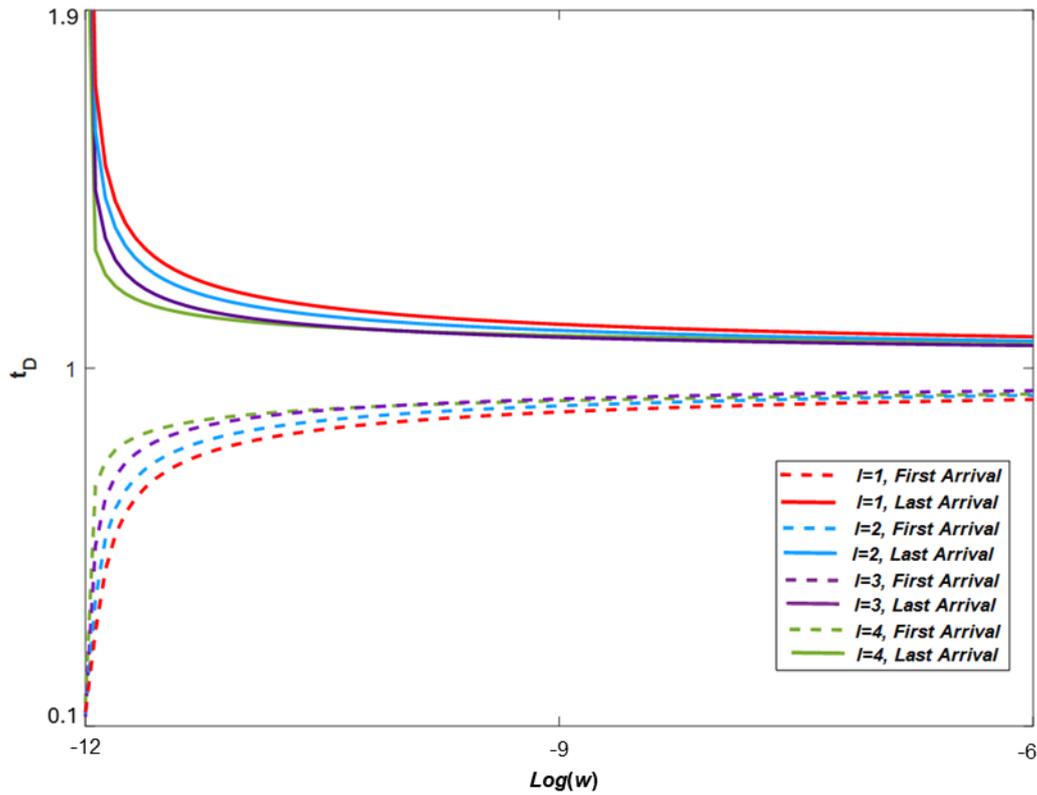

Figure 5. Effect of water velocity ($w$) and exponent $l$ on the first and last arrival times.

Figure 6 presents the effect of water velocity direction on $CO_2$ plume behaviour by showing the positions of the initial and rear fronts at different times and water velocities. To quantify this influence, the ratio between the water velocity ($w$) and the velocity of the moving $CO_2$ plume ($u$) is used. Positive values of this ratio indicate upward water flow in the same direction as the migrating gas, while negative values correspond to downward water flow. For clearer comparison, the case with no water velocity—representing a scenario with stagnant water—is also included, denoted by the dashed curve. This provides a direct comparison of how water movement influences plume dynamics relative to the baseline scenario.

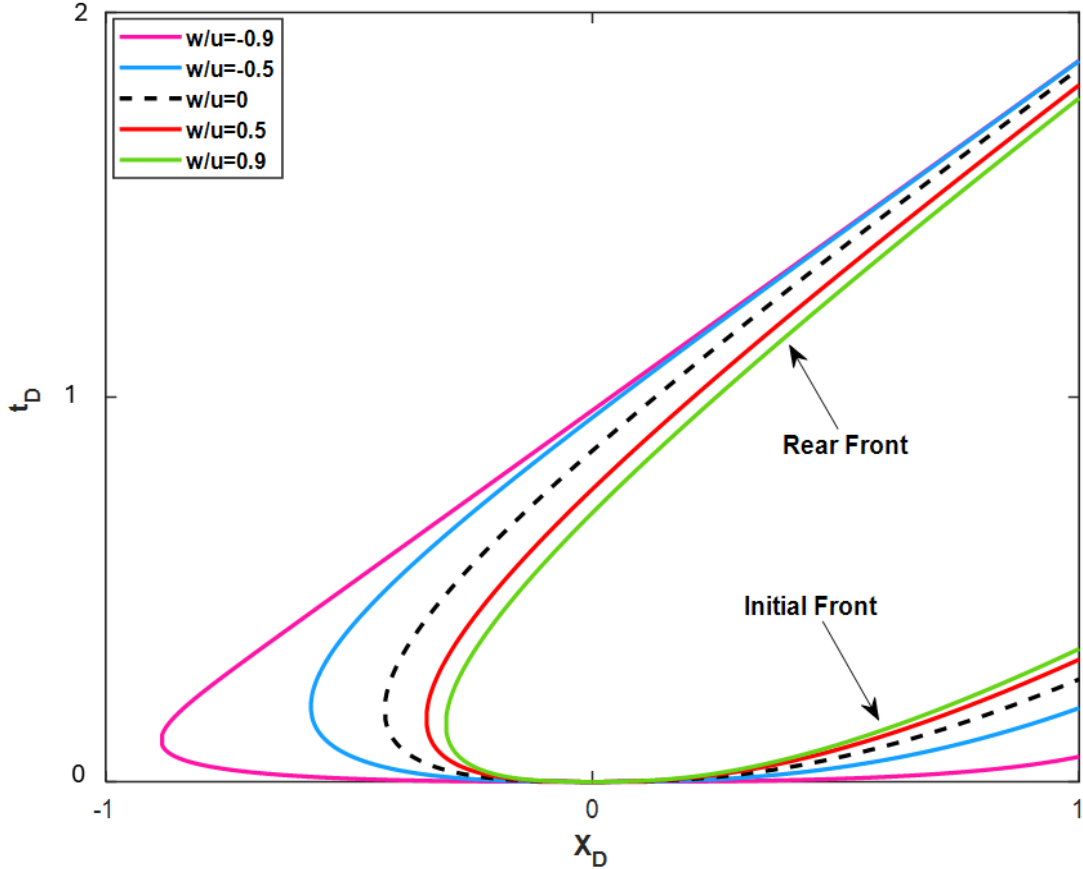

Figure 6. Effect of water velocity direction on the initial and rear front trajectories.

As evident from Figure 6, the direction of water flow significantly influences the accumulation period, defined as the difference between the first and last arrival times. When the direction of water flow changes from upward (positive values) to downward (negative values), the accumulation period notably increases. Moreover, increasing the water velocity—whether upward or downward—intensifies this effect.

The underlying mechanism behind this behaviour is that upward water flow assists in the migration of the $CO_2$ plume by exerting an external pressure that pushes the plume upwards. This upward pressure enhances the buoyancy force, facilitating the plume's movement towards the storage trap and consequently reducing the accumulation period. In contrast, downward water flow creates a resistance to the plume's movement, impeding its progress and increasing the time required for the $CO_2$ to accumulate in the trap. Thus, the dynamics of water flow, whether aiding or resisting the plume, play a crucial role in determining the duration of $CO_2$ storage within a reservoir.

Another key observation from this analysis is that, regardless of the direction of water flow, the velocity of the initial front always remains positive. In contrast, the rear front initially moves down-dip before reversing direction and moving up-dip, even when there is no water flow. This indicates that, while water flow has a clear influence on plume movement, the rear front's motion is also governed by other factors such as buoyancy, reservoir characteristics, and the nature of the seal, which cause the rear front to initially move down-dip before ultimately progressing upwards toward the trap. This highlights the complex interactions that occur between the $CO_2$ plume and the reservoir during the injection and storage process.

3.2 Effect of chemical reaction coefficient on $CO_2$ plume behaviour

Another important parameter that influences the behaviour of the $CO_2$ plume is the chemical reaction coefficient ($k_a$). This coefficient represents the rate at which $CO_2$ interacts with the surrounding rock and fluids, affecting the reservoir's properties. A higher chemical reaction coefficient accelerates the dissolution of $CO_2$ in water, leading to the formation of carbonic acid, which can alter the reservoir's mineralogy. These changes can enhance or hinder $CO_2$ migration, impacting the plume's dynamics, its rate of propagation, and the long-term stability of $CO_2$ storage.

Figure 7 illustrates the influence of the chemical reaction coefficient ($k_a$) on both the first and last arrival times. As depicted in the figure, for a fixed $l$ value, an increase in the chemical reaction coefficient leads to a delay in the first arrival time and a reduction in the last arrival time. This behaviour can be attributed to the impact of chemical reactions on the migration of $CO_2$ through the reservoir. Specifically, as $k_a$ increases, the rate of chemical reactions between the migrating $CO_2$, the formation water, and the reservoir rock intensifies. This results in a deceleration of the first front of the plume, causing the $CO_2$ to take longer to reach the trap. The enhanced chemical reactivity effectively "retards" the migration of $CO_2$, as it is either dissolved into the formation water or converted into new carbonate minerals, thereby reducing its mobility. These processes contribute positively to the trapping mechanisms of $CO_2$, helping to secure the gas within the reservoir over the long term.

However, while the increased chemical reactions slow down the first arrival time, they simultaneously reduce the total volume of migrating $CO_2$. This reduction in $CO_2$ volume occurs due to the consumption of $CO_2$ during the chemical reactions, which leads to a decrease in the amount of $CO_2$ available to propagate through the reservoir. As a result, the last arrival time (corresponding to the rear front of the plume) is shortened, accelerating the end of the accumulation period. This dual effect—slowing the initial front while accelerating the end of accumulation period—highlights the complex interplay between chemical reactions and $CO_2$ migration dynamics in the context of underground storage.

Another important observation from Figure 7 is that for a chemical reaction coefficient greater than 2500, the first and last arrival times coincide, indicating that the accumulation period is zero. This can be explained by noting that, as $k_a$ increases, the rate of chemical reactions between the injected $CO_2$ and the surrounding rock and fluids also increases. As a result, the entire mass of injected $CO_2$ is consumed due to these chemical reactions, leading to the complete sequestration of $CO_2$ before it reaches the trap.

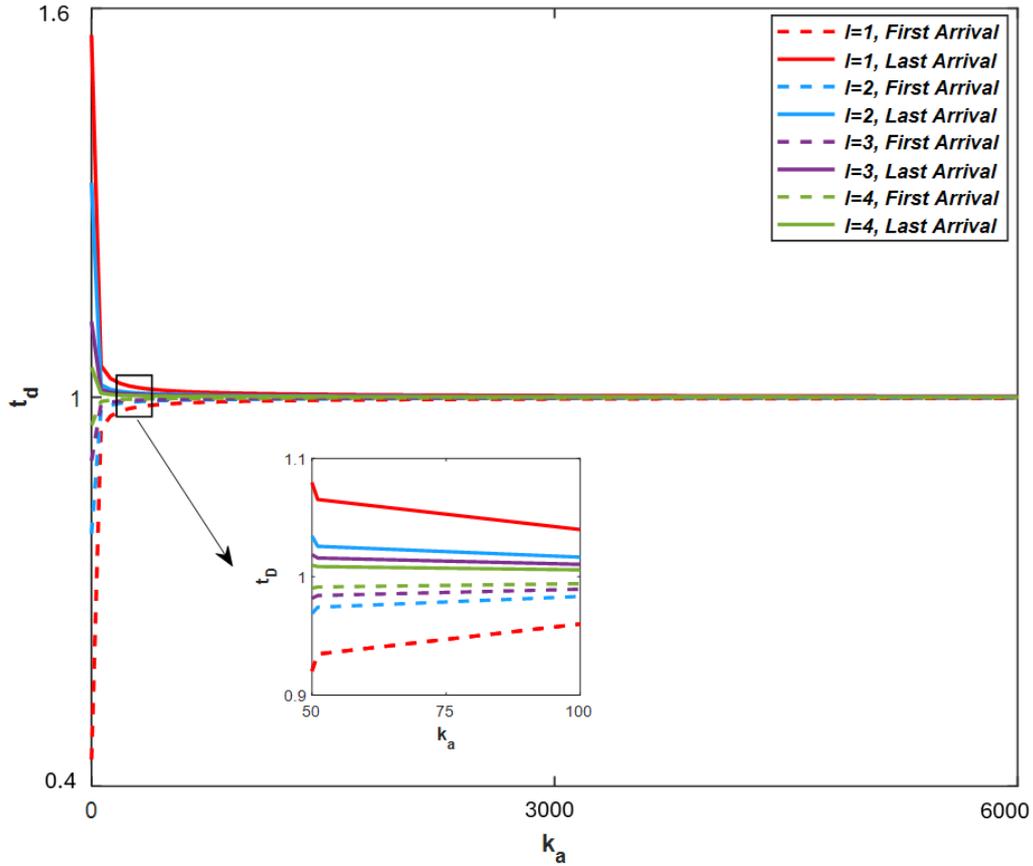

Figure 7. Effect of chemical reaction coefficient $k_a$ and exponent $l$ on the first and last arrival times.

Figure 8 illustrates the effect of the chemical reaction coefficient ($k_a$) on the trajectories of the initial and rear fronts at different times, supporting the observations in Figure 7. While higher $k_a$ generally slows $CO_2$ migration towards the trap, it also reduces the accumulation period due to sequestration through chemical reactions. This process enhances long-term containment by converting $CO_2$ into stable forms, thereby reducing the volume of mobile $CO_2$ in the reservoir. These findings highlight the need to account for both migration dynamics and chemical reactions when assessing $CO_2$ storage effectiveness.

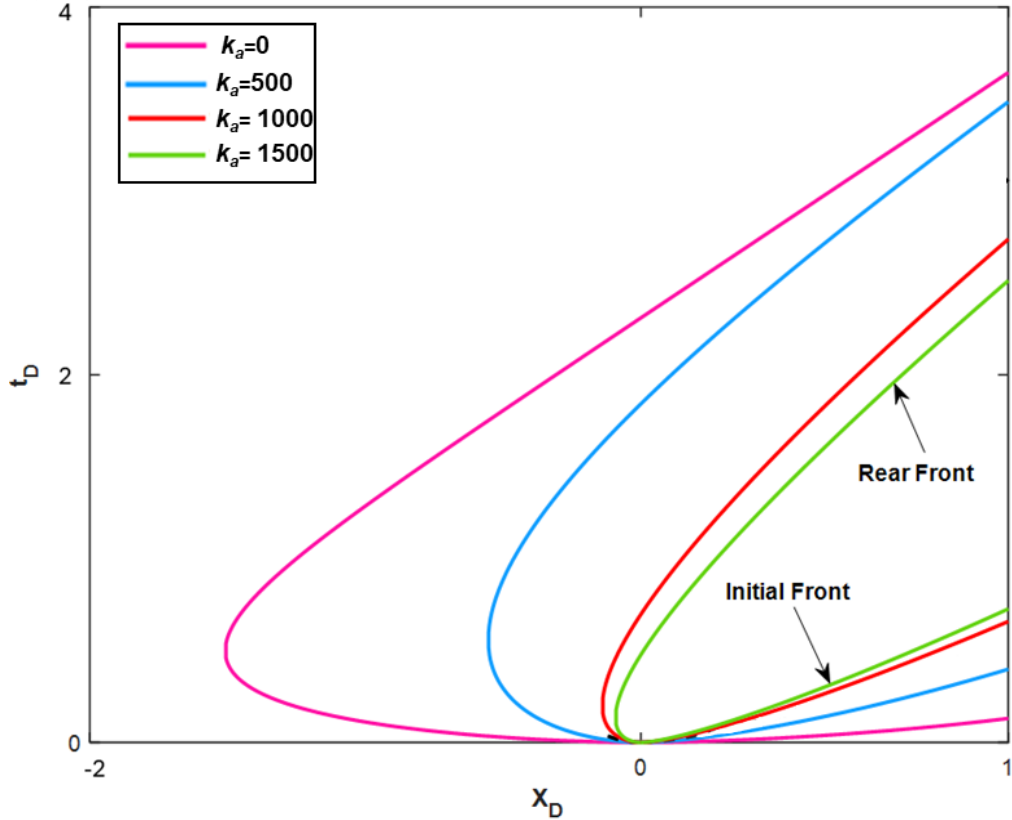

Figure 8. Effect of chemical reaction coefficient $k_a$ on the initial and rear front trajectories.

## 3.3 Position of $CO_2$ plume

To illustrate the application of the model in a practical context, it is assumed that the volume of the trap is negligible in comparison to the injected volume for the basic case described above. The position of the $CO_2$ plume, $h(x,t)$, is calculated after 100, 200, and 300 years. Figure 9 displays the $CO_2$ plume position at these different time intervals. As shown in Figure 9, as time progresses, the injected $CO_2$ spreads beneath the sealing surface, with its height decreasing over time. Additionally, due to the incorporation of chemical reactions in the model, the mass of the migrating $CO_2$, represented by the area under each curve, decreases as it undergoes transformation through these reactions.

It is important to note that the pulse injection model is employed for this calculation, as the injection period is shorter than the migration period of $CO_2$ within the reservoir in this particular scenario.

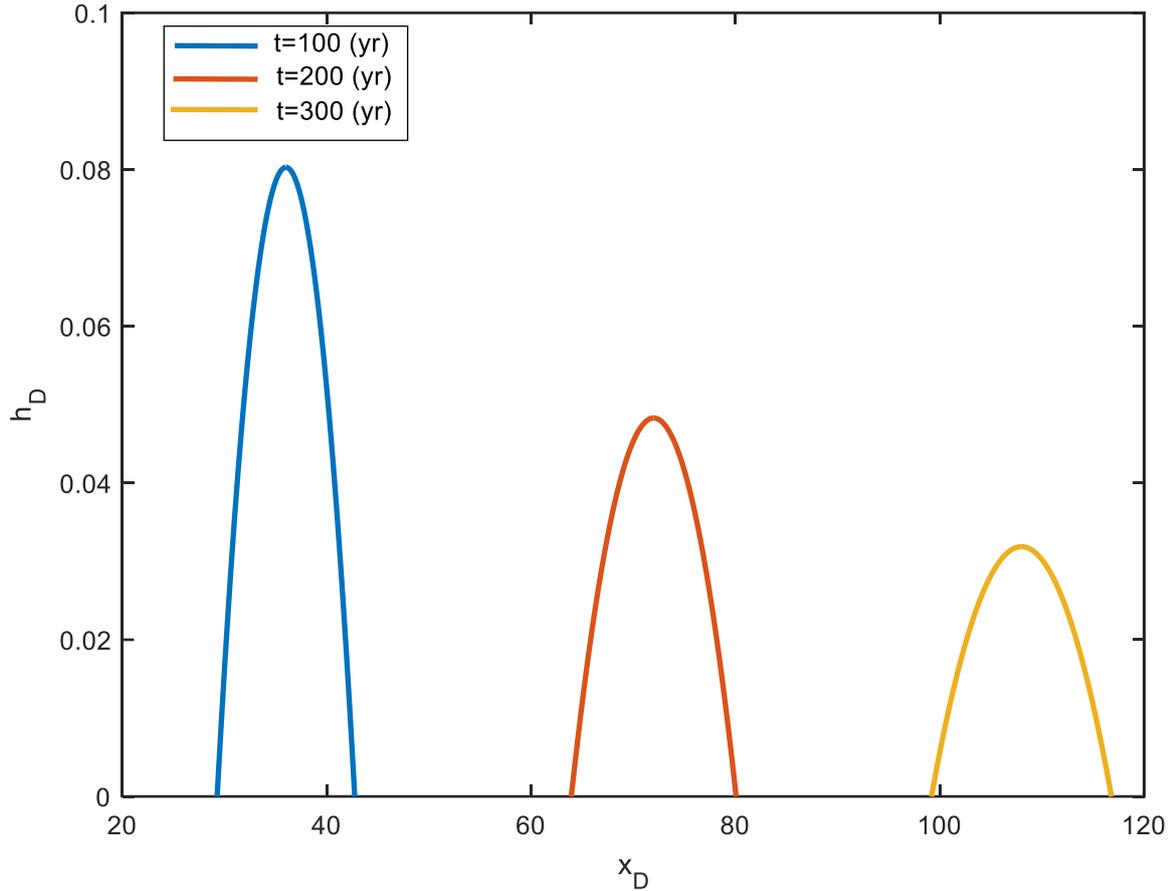

Figure 9. The position of the gas plume after 100, 200, and 300 years.

## 4 Discussion

This study provides exact solutions for predicting the dynamics, propagation, and dispersion of the injected gas plume, along with determining its boundaries beneath the seals during subsurface reservoir storage. The dynamic equations for gravity-driven flows are derived, and the corresponding self-similar solutions are subsequently determined. The solution for pulse injection describes the long-term evolution of the plume, whereas the solution for continuous injection introduces a time scale that aids in understanding long-term propagation.

The model for pulse injection assumes the instantaneous injection of a finite mass at $t=0$, but in reality, the injection occurs over a finite duration, denoted by $T$. This assumption is reflected in equation (27), where at $t=0$, the height of the plume tends towards infinity, a result that lacks physical relevance. To address this issue, the injection period $T$ is determined by equating the first moments of the solution $h(x,t)$ for pulse injection. Once $T$ is calculated, the plume's behaviour during this period will approximate that of a continuous injection. Therefore, the solution for continuous injection, presented in equation (34), can be applied to model the plume's evolution during this period. After $T$, the exact solution for pulse injection, as provided in equation (27), can be used to predict the plume's behaviour. This approach effectively resolves the singularity at $t=0$ in the pulse injection model and ensures a physically meaningful representation of the plume dynamics.

Regarding the power-law shape of the stringers' cross-sectional area, the exponent $l$ determines the shape of the cross-section. As $l$ increases from values less than one to values greater than one, the cross-sectional area available for the flow also increases, causing the shape of the cross-section to transition from sharp to flatter (see Figure 2). With an increase in the stringers' cross-sectional area—resulting from an increase in $l$ —the area available for the flowing injected gas expands. As a result, for the same volumetric flux, the velocity of the gas plume decreases. This is why the arrival time increases for larger values of $l$, as shown in Figures 5 and 7. In the case where all the stringers are filled with gas, and the gas height exceeds the height of the stringers, the flow becomes linear. In this scenario, $l$ should be set equal to one to reflect this effect.

In terms of improving the model, several key areas could enhance the accuracy of the derived solutions. In this study, the cap rock is treated as a smooth layer, whereas in reality, the seal is heterogeneous, containing spill points and intermediate traps. This represents an area for model improvement in future studies. Additionally, both gas and water are treated as incompressible fluids, with constant fluid densities assumed throughout the process. Incorporating compressibility and allowing for variations in fluid densities would yield more accurate results, though it would increase the model's complexity.

# 5   Conclusions

This study examined the analytical modelling of segregated flow, incorporating the effects of chemical reactions and water counter-flow, to enhance the understanding of sssn porous media. The key findings of our investigation can be summarised as follows:

1. An increase in the chemical reaction coefficient ($k_a$) leads to a delay in the first arrival time, while reducing the last arrival time. This indicates that chemical reactions, particularly dissolution and mineralization processes, can slow the migration of $CO_2$, enhancing long-term sequestration.

2. Under the absence of water counter-flow, the dispersive flux is negligibly lower than the advective flux. Water counter-flow significantly increases the dispersive flux, which can exceed the advective flux.

3. An increase in down-dip water velocity accelerates the leading front of the $CO_2$ plume while decelerating the rear front. Conversely, reversing the water flow direction from down-dip to up-dip shortens the accumulation period, thereby reducing the time required for $CO_2$ to be stored within the geological trap. This effect significantly influences the migration and accumulation dynamics of the $CO_2$ plume.

4. The power-law shape of the stringers' cross-sectional area affects the plume's dynamics, where an increase in the exponent $l$ expands the cross-sectional area, thereby reducing the velocity of the gas plume and increasing the arrival time.

Future model improvements could include accounting for heterogeneity in the cap rock, as well as incorporating compressibility and variable fluid densities, which would provide more accurate predictions while increasing the model's complexity. These findings underscore the importance of understanding chemical reactions and water flow dynamics in governing $CO_2$ migration and storage efficiency. This knowledge is crucial for optimising $CO_2$ storage strategies and ensuring the long-term stability of carbon capture and storage initiatives in aquifers.

**References**

Barenblatt GI (2003) 'Scaling.' (Cambridge University Press).

Bedrikovetsky P, Borazjani S (2022) Exact solutions for gravity- segregated flows in porous media. *Mathematics* **10**(14), 2455. doi:10.3390/math10142455.

Bjørlykke K (1993). Fluid flow in sedimentary basins. *Sedimentary Geology*, *86*(1-2), 137-158.

Carroll, S., Carey, J.W., Dzombak, D., Huerta, N.J., Li, L., Richard, T., Um, W., Walsh, S.D. and Zhang, L., 2016. Role of chemistry, mechanics, and transport on well integrity in CO2 storage environments. *International Journal of Greenhouse Gas Control*, *49*, pp.149-160.

Ciriello V, Di Federico V, Archetti R, Longo S (2013) Effect of variable permeability on the propagation of thin gravity currents in porous media. *International Journal of Non-Linear Mechanics* **57**, 168–175. doi: 10.1016/j.ijnonlinmec.2013.07.003.

Dentz M, Tartakovsky DM (2009) Abrupt-interface solution for carbon dioxide injection into porous media. *Transport in Porous Media* **79**(1), 15–27. doi:10.1007/s11242-008-9268-y.

Elizabeth TP, Valiya MH (2014). Determination of fluid flow in Deep Sedimentary Layers in the Campos Basin. In *VI Brazilian Geophysics Symposium, Porto Alegre*, *Rio Grande do Sul, Brazil*.

Fahs M, Younes A, Mara TA (2014) A new benchmark semi-analytical solution for density-driven flow in porous media. *Advances in Water Resources* **70**, 24–35. doi: 10.1016/j.advwatres.2014.04.013.

Golding MJ, Huppert HE (2010) The effect of confining impermeable boundaries on gravity currents in a porous medium. *Journal of Fluid Mechanics* **649**, 1–17. doi:10.1017/S0022112009993223.

Hantschel T, Kauerauf AI (2009). Fundamentals of basin and petroleum systems modeling. Springer Science & Business Media.

Huppert HE, Woods AW (1995) Gravity-driven flows in porous layers. *Journal of Fluid Mechanics* **292**, 55–69. doi:10.1017/S0022112095001431.

Longo S, Ciriello V, Chiapponi L, Di Federico V (2015) Combined effect of rheology and confining boundaries on spreading of gravity currents in porous media. *Advances in Water Resources* **79**, 140–152. doi: 10.1016/j.advwatres.2015.02.016.

Lyle S, Huppert HE, Hallworth M, Bickle M, Chadwick A (2005) Axisymmetric gravity currents in a porous medium. *Journal of Fluid Mechanics* **543**, 293–302. doi:10.1017/S0022112005006713.

Pegler SS, Huppert HE, Neufeld JA (2013) Topographic controls on gravity currents in porous media. *Journal of Fluid Mechanics* **734**, 317–337. doi:10.1017/jfm.2013.466.

Rochelle, C. A., Czernichowski-Lauriol, I., & Milodowski, A. E. (2004). The impact of chemical reactions on CO2 storage in geological formations: a brief review. *Geological Society, London, Special Publications*, *233*(1), 87-106.

Tosco T, Petrangeli Papini M, Cruz Viggi C, Sethi R (2014) Nanoscale zerovalent iron particles for groundwater remediation: A review. *Journal of Cleaner Production* **77**, 10–21. doi: 10.1016/j.jclepro.2013.12.026.

Vella D, Huppert HE (2006) Gravity currents in a porous medium at an inclined plane. *Journal of Fluid Mechanics* **555**, 353–362. doi:10.1017/S0022112006009578.

Williams GA, Chadwick RA, Vosper H (2018) Some thoughts on darcy- type flow simulation for modelling underground CO2 storage, based on the sleipner CO2 storage operation. *International Journal of Greenhouse Gas Control* **68**, 164–175. doi: 10.1016/j.ijggc.2017.11.010.

Zheng Z, Soh B, Huppert HE, Stone HA (2013) Fluid drainage from the edge of a porous reservoir. *Journal of Fluid Mechanics* **718**, 558–568. doi:10.1017/jfm.2012.630.

Zheng Z, Christov IC, Stone HA (2014) Influence of heterogeneity on second-kind self-similar solutions for viscous gravity currents. *Journal of Fluid Mechanics* **747**, 218–246. doi:10.1017/jfm.2014.148.

Zheng Z, Guo B, Christov IC, Celia MA, Stone HA (2015) Flow regimes for fluid injection into a confined porous medium. *Journal of Fluid Mechanics* **767**, 881–909. doi:10.1017/jfm.2015.68.